\documentclass[pdflatex]{article} 

\usepackage{url}
\usepackage[onehalfspacing]{setspace}
\usepackage{amssymb,amsfonts,amsmath}
\usepackage[pdftex]{graphicx,color}
\usepackage{natbib}
\usepackage[margin=1in]{geometry}

\newcommand{\pdiff}[2]{\frac{\partial #1}{\partial #2}}
\newcommand{\di}{\mathrm{d}}
\newcommand{\xv}{\boldsymbol{x}}
\newcommand{\yv}{\boldsymbol{y}}
\newcommand{\zv}{\boldsymbol{z}}
\newcommand{\uv}{\boldsymbol{u}}
\newcommand{\Am}{\boldsymbol{A}}
\newcommand{\Bm}{\boldsymbol{B}}
\newcommand{\Cm}{\boldsymbol{C}}
\newcommand{\Dm}{\boldsymbol{D}}
\newcommand{\Em}{\boldsymbol{I}}
\newcommand{\Rm}{\boldsymbol{R}}
\newcommand{\Um}{\boldsymbol{U}}

\newcommand{\Xm}{\boldsymbol{X}}
\newcommand{\Km}{\boldsymbol{K}}
\newcommand{\Ym}{\boldsymbol{Y}}
\newcommand{\Fm}{\boldsymbol{F}}
\newcommand{\Gm}{\boldsymbol{G}}
\newcommand{\Hm}{\boldsymbol{H}}
\newcommand{\Zm}{\boldsymbol{Z}}
\newcommand{\onev}{\boldsymbol{1}}
\newcommand{\gv}{\boldsymbol{g}}
\newcommand{\ev}{\boldsymbol{e}}
\newcommand{\vv}{\boldsymbol{v}}
\newcommand{\wv}{\boldsymbol{w}}
\newcommand{\av}{\boldsymbol{a}}
\newcommand{\tv}{\boldsymbol{t}}
\newcommand{\phiv}{\boldsymbol{\phi}}
\newcommand{\Phim}{\boldsymbol{\Phi}}
\newcommand{\im}{\mathrm{i}}
\newcommand{\trans}[1]{{#1}^{T}}
\newcommand{\inv}[1]{{#1}^{-1}}

\DeclareMathOperator{\diag}{diag}
\DeclareMathOperator{\erf}{erf}
\DeclareMathOperator{\Tr}{Tr}
\newcommand{\vd}{\boldsymbol{\delta}}

\begin{document}
\begin{center}
  \Large
  The most probable neural circuit exhibits low-dimensional sustained activity

  \vspace{1cm}
  
  \large
  Takuma Tanaka

  Faculty of Data Science, Shiga University, Hikone, 522-8522, Japan
  
  tanaka.takuma@gmail.com
\end{center}

\section*{Abstract}
Cortical neurons whose activity is recorded in behavioral experiments has been classified into several types such as stimulus-related neurons, delay-period neurons, and reward-related neurons.
Moreover, the population activity of neurons during a reaching task can be described by dynamics in low-dimensional space.
These results suggest that the low-dimensional dynamics of a few types of neurons emerge in the cerebral cortex that is trained to perform a task.
This study investigates a simple neural circuit emerges with a few types of neurons.
Assuming an infinite number of neurons in the circuit with connection weights drawn from a Gaussian distribution, we model the dynamics of neurons by using a kernel function.
Given that the system is infinitely large, almost all capable circuits approximate the most probable circuit for the task.
We simulate two delayed response tasks and a motor-pattern generation task.
The model network exhibits low-dimensional dynamics in all tasks.
Considering that the connection weights are drawn from a Gaussian distribution, the most probable circuit is the circuit with the smallest connection weight; that is, the simplest circuit with the simplest dynamics.
Finally, we relate the dynamics to algorithmic information theory.

\section{Introduction}
The activity of neurons has been recorded from animals performing behavioral tasks, most of which include delay periods.
The delay periods before and between motor execution are accompanied by sustained neuronal activity in the motor-related cortical areas \citep{Funahashi1989,Shima2000}.
Sustained activity is ubiquitously observed in the unit recordings of neocortical neurons \citep{Niki1974,Fuster1982,Latimer2015}.
\cite{Shima2000} reported that 317 out of 1027 SMA and pre-SMA task-related neurons exhibit activity between two movements; \cite{Barone1989} found sustained activity in 35 out of 302 prefrontal neurons following a stimulus presentation.
Sustained activity is thought to encode previously presented stimuli and the planned action in delayed response tasks and other tasks requiring short-term memory.

Therefore, the sustained activity of a large number of neurons in an area is thought to represent memory stored in the brain.
However, this notion is less natural than it seems.
Given the analogous workings of computer programs and brain processing, let us consider a computer program performing a delayed response task.
The computer program receives a sequence of inputs from an input device and sends a sequence of outputs to an output device.
The input--output relationship is determined by the rules of the task,
and the input sequence and generated output sequence are stored as variables
in the memory space.
If the number of input sequences used in the task is comparable to that used in behavioral tasks, the input sequence is represented by a variable with several bits.
Considering that modern computers typically contain several gigabytes of semiconductor memory, the fraction of semiconductors encoding the input sequences is negligible.
Therefore, if a probe is randomly inserted into the main memory of a computer, the semiconductors encoding the input and output sequences will be found with extremely low probability.
By contrast, the neurons encoding the previous stimulus and a planned action are easily found among the motor cortical neurons of animals that are performing tasks.
The representation is far more redundant in the brain than in computers.

Equally impressive is the difference between animals generating a motor pattern and computers performing a similar task.
\cite{Churchland2012} reported the activity of motor cortical neurons in monkeys performing several distinct arm movements.
The recorded multineuron activity obeyed low-dimensional dynamics with the same rotational component during different arm movements.
As in delayed response tasks, these low-dimensional dynamics are absent in the semiconductors of a computer simulating similar motor patterns.

To understand the origin of the different behaviors of cortical neurons and computer programs running on semiconductors,
this article develops a mathematical framework with the following features.
First, the neural networks are constructed from an infinite number of neurons.
Second, in this limit case, we derive the network structure's probability, which is independent of the learning rule.
Third, we show that the most probable network exhibits the simplest structure, which corresponds to the shortest program in algorithmic information theory.

This remainder of this article is organized as follows.
Section 2 states the hypothesis and formulates the mathematical model.
The building block of the network is a discrete-time binary neuron model.
Under the presented assumptions, we estimate the probability of forming a network structure by summing the squared connection weights.
In the limit of infinitely many neurons, most of the network structures that can perform the task are concentrated around the most probable network structure.
Hence, the most probable network structure is a typical network structure.
In this limit case, we derive a continuous-value model and its kernel representation.
Section 3 presents the simulation results of several tasks (an AND task, a delayed response task, and a motor-pattern generation task).
In these simulations, the model networks exhibit sustained activity in the delay periods.
The dynamics of the sustained activity are qualitatively similar to experimental results.
Section 4 summarizes the results and discusses their relationship to algorithmic information theory, along with the limitations and potential extensions of the model.

\section{Model}


\subsection{Model neuron and model network}
Consider a model network of $N_0$ neurons, which receive input from other neurons and generate continuous-valued output.
The output $x_i^t$ of neuron $i$ at time step $t$ is defined by

\begin{equation}
x_i^t = f\left(\sum_{j=1}^{N_0} W_{ij} x_j^{t-1}-h_i\right),
\end{equation}

where
$x_i^t$ is the state of neuron $i$ at time step $t$,
$W_{ij}$ is the connection weight from neuron $j$ to $i$,
$h_i$ is the threshold of neuron $i$, and
the function $f$ determines the firing probability.
Here we assume that $f$ is a sigmoidal function.
$h_i$ can be any real value.
Self-connection is allowed; that is, $W_{ii}$ can be a nonzero value.
In this model, a single neuron can make both excitatory (positive) and inhibitory (negative) connections to other neurons.
Let us assume that $W_{ij}$ can be $-w N_0^{-1/2}$ and $w N_0^{-1/2}$ with equal probability, where $w$ is a positive constant.
The factor $N_0^{-1/2}$ ensures that the variance of $\sum_{j=1}^{N_0} W_{ij}$ is invariant with $N_0$.
The connection weight matrix has $2^{N_0^2}$ possible configurations.

The network configuration determines the ability of the network to perform a given task.
Let us estimate the probability of a configuration that permits the task accomplishment.
This is equivalent to integrating the probability over the network configurations that can perform the task.
To estimate this probability, we assume an infinite number of neurons in the network and coarse-grain the network.
By averaging the state of infinitely many stochastic binary neurons with the same connection weights, we obtain a deterministic sigmoid neuron
with the following dynamics:

\begin{equation}
x_i^t = f\left(\sum_{j=1}^{N_0} W_{ij} x_j^{t-1}-h_i\right),
\end{equation}
where $N_0$ is redefined as the number of sigmoid neurons.

\begin{figure}
  \centering
  \includegraphics{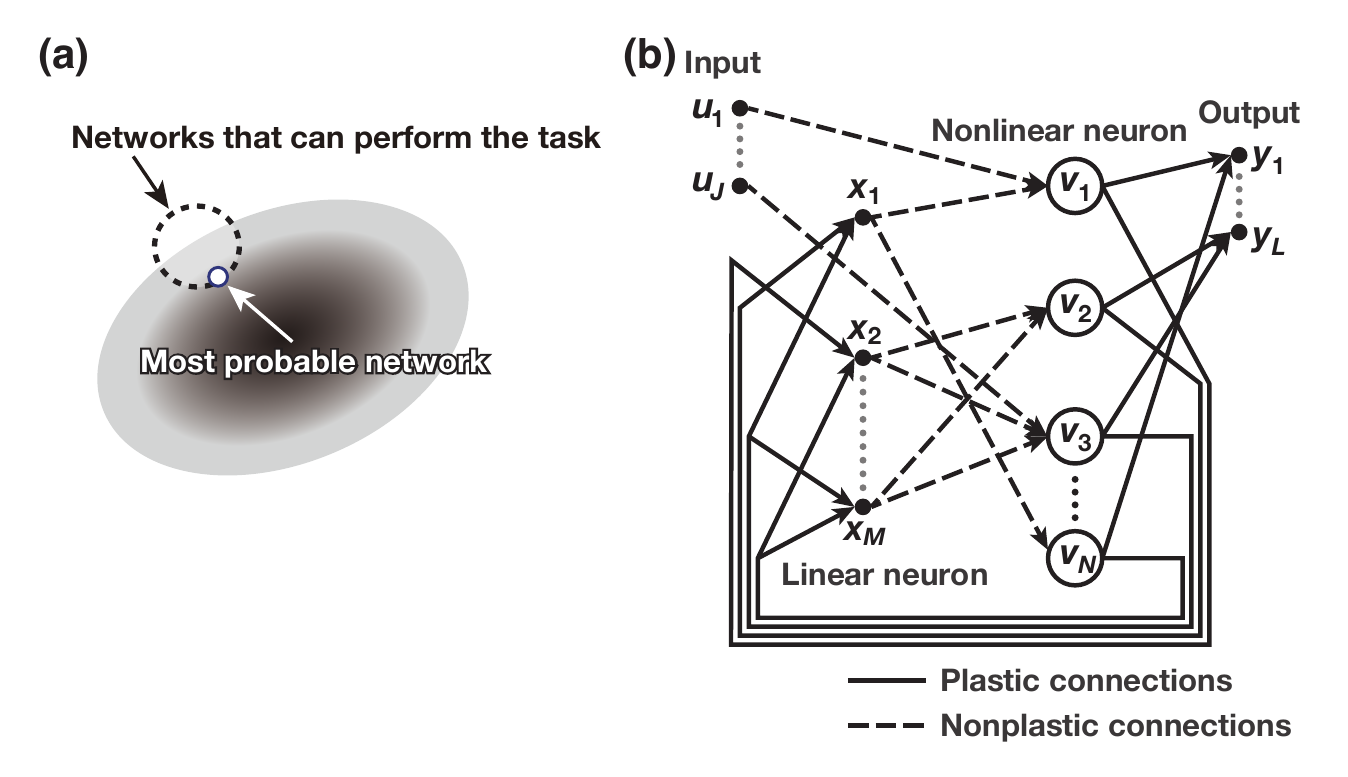}
    \caption{\label{schema}
      (a) In the space of possible network structures, the dashed circle indicates the networks that can perform the task.
      The dark gray area is the region of high probability density.
      (b) The model network is composed of input, output, linear, and nonlinear neurons. The plastic and nonplastic connections are indicated by solid and dashed lines, respectively.
    }
\end{figure}

The probability that the summed connection weights from a group of $n_j$ neurons received by a target neuron exceeds $W_{\mathrm{bound}}$ is approximated by

\begin{equation}
  p \approx \int_{W_{\mathrm{bound}}}^\infty\sqrt{\frac{N_0}{2\pi n_j w^2}}\exp\left(-\frac{N_0x^2}{2n_j w^2}\right)\di x,
\end{equation}
where the binomial distribution is approximated by the Gaussian distribution.
The probability that the same situation holds for $n_i$ neurons is approximated by $p^{n_i}$.
In the limit of $N_0\rightarrow\infty$, the log-probability is approximated by

\begin{equation}
  \log p^{n_i} \approx -\frac{N_0 n_i W_{\mathrm{bound}}^2}{2 n_j w^2} \propto -N_0\frac{n_i}{n_j}W_{ij}^2, \label{wapprox}
\end{equation}
where we have used the saddle-point method with $W_{ij}$ redefined by $W_{\mathrm{bound}}$.
Equation~(\ref{wapprox}) defines the probability that the capable network structures for the given task will appear, provided that the connection weights of the neurons in group $j$ to a neuron in group $i$ exceeds $W_{ij}$ to perform the task.
If the total connection weight received by each neuron in group $i$ exceeds $W_{ij}$ or is below $-W_{ij}$, the logarithm of the realization probability of the network is proportional to
\begin{equation}
  -N_0\sum_{i,j=1}^{N_G} \frac{n_i}{n_j}W_{ij}^2 \equiv -\Lambda_0, \label{probability1}
\end{equation}
where $n_i$ is the number of neurons in group $i$ and $N_G$ is the number of groups.
To obtain the most probable network structure that can perform a task, we can maximize $-\Lambda_0$ with respect to $n_i$ and $W_{ij}$, provided that the network structure can perform the task (Figure~\ref{schema}a).

Note that partitioning group $i$ into subgroups $i'$ and $i''$ does not affect the approximated log-probability because

\begin{align}
& -N_0\frac{n_i}{n_j}W_{ij}^2-N_0\frac{n_j}{n_i}W_{ji}^2 \nonumber\\
= & 
-N_0\frac{n_{i'}}{n_j}W_{i'j}^2-N_0\frac{n_j}{n_{i'}}W_{ji'}^2
-N_0\frac{n_{i''}}{n_j}W_{i''j}^2-N_0\frac{n_j}{n_{i''}}W_{ji''}^2
\end{align}
if

\begin{align}
  n_i=&n_{i'}+n_{i''} \label{con0},\\
  W_{ij}=&W_{i'j}=W_{i''j}, \label{con1}\\
  W_{ji}=&W_{ji'}+W_{ji''}, \label{con1b}\\
\frac{W_{ji}}{n}=&\frac{W_{ji'}}{n_{i'}}=\frac{W_{ji''}}{n_{i''}}. \label{con2}
\end{align}
Also note that Equation~(\ref{con2}) maximizes the probability under Equations~(\ref{con0}) and (\ref{con1b}).
This means that the groups can be partitioned and merged without affecting the most probable network structure.

There are $\frac{N_0!}{\prod_{i=1}^{N_G} n_i!}$ possible assignments of $N_0=\sum_{i=1}^{N_G} n_i$ neurons into $N_G$ groups.
The logarithm of this number of assignments is approximately $-\sum_{i=1}^{N_G} n_i\log\frac{n_i}{N_0}$, which is of order $N_0$.
This term is maximized if $n_i=N_0/N_G$.
Thus, by assuming $n_i=N_0/N_G$ in Equation~(\ref{probability1}), we minimize

\begin{equation}
\Lambda_0 = \sum_{i,j=1}^{N_G} W_{ij}^2
\end{equation}
given that the network structure can perform the task.

\subsection{Representing the model by the kernel method}
The saddle-point method is justified if the network contains a sufficiently large number of neurons.
To simplify the previous model with an infinite $N_G$, we divide the network into two populations: one containing linear neurons and the other containing nonlinear neurons (Figure~\ref{schema}b).
The linear and output neurons receive inputs from input neurons and nonlinear neurons.
Nonlinear neurons receive inputs from input neurons and linear neurons.
The network dynamics are described by

\begin{align}
v_i^{t+1} =& f\left(\sum_{j=1}^J W_{ij}^{u\rightarrow v} u_j^t+\sum_{j=1}^M W_{ij}^{x\rightarrow v} x_j^t
-h_i\right),\\
x_i^t =& \sum_{j=1}^N W_{ij}^{v\rightarrow x} v_i^t,\\
y_i^t =& \sum_{j=1}^N W_{ij}^{v\rightarrow y} v_i^t,
\end{align}
where $u_i$, $x_i$, $v_i$, and $y_i$ are the states of the input, linear, nonlinear, and output neurons, respectively.
Assume that the network contains $J$ input, $M$ linear, $N$ nonlinear, and $L$ output neurons.
Assume also that the connection weights from input and linear neurons to nonlinear neurons ($W_{ij}^{u\rightarrow v}$ and $W_{ij}^{x\rightarrow v}$ respectively) are constants drawn from a Gaussian distribution.
Only the connection weights to the linear and output neurons ($W_{ij}^{v\rightarrow x}$ and $W_{ij}^{v\rightarrow y}$ respectively) are plastic, that is, modifiable.
Consider that this model contains $M+N+L$ groups of neurons with some activation functions $f(x)=x$ and some connection weights set to zero.
The most probable network structure is the network structure with the minimal $L^2$ norm of connection weights to the linear and output neurons.

Assuming an infinite number of nonlinear neurons, that is, $N\rightarrow\infty$, we can use the kernel trick.
By setting $f(x)=\erf(x)$ and assuming that the connection weights from linear to nonlinear neurons are drawn from an isotropic Gaussian distribution with mean zero and variance $\sigma_0^2/N$, and that the bias terms are drawn from an isotropic Gaussian distribution with mean zero and variance $\sigma_b^2$, the kernel function is given by

\begin{equation}
K_{tt'} = \frac{2}{\pi}\arcsin\left(2\frac{\sigma_0^2 Z_{tt'}+\sigma_b^2}{
\sqrt{[(1+2\sigma_b^2)+2\sigma_0^2 Z_{tt}][(1+2\sigma_b^2)+2\sigma_0^2 Z_{t't'}]}
}\right)
\end{equation}
\citep{Williams1997,Hermans2012}, where

\begin{align}
F_{tt'} = & \alpha\Em+\sum_{j=1}^J u_j^t u_j^{t'},\\
G_{tt'} = & \sum_{m=1}^M x_m^t x_m^{t'},\\
H_{tt'} = & \sum_{l=1}^L y_l^t y_l^{t'},\\
Z_{tt'} = & F_{tt'}+G_{tt'}.
\end{align}
This kernel function is obtained by integrating over all possible connection weights.
$\alpha$ distinguishes the inputs at time $t$ from those at time $t'$ even when $\uv^t=\uv^{t'}$.
In this paper, $\alpha$ is set to $10^{-6}$.
We assume periodic boundary conditions in the time domain, that is, $\uv^{T+1}=\uv^1$, $\xv^{T+1}=\xv^1$, and $\yv^{T+1}=\yv^1$, where $T$ is the duration of the simulated task.
As shown in Appendix~\ref{L2norm}, the $L^2$ norm of the connection weights is given by

\begin{equation}
\Lambda = \Tr[\inv\Km(\Gm^+ +\Hm^+)]. \label{V}
\end{equation}
For $T\times T$ matrix $\Am$,  $\Am^+$ is given by

\begin{equation}
  (\Am^+)_{tt'} = \begin{cases}
    A_{t+1\; t'+1} & t<T,\; t'<T\\
    A_{1\;t'+1} & t=T,\; t'<T\\
    A_{t+1\;1} & t<T,\; t'=T\\
    A_{11} & t=T,\; t'=T
  \end{cases}.
\end{equation}

Although we have assumed that infinitely many linear neurons, we can set $M$ to a finite number because the objective function [Equation~(\ref{V})] depends only on the Gram matrices of the input, linear, and nonlinear neurons.
The $T\times T$ Gram matrix $\Gm$ is the summation of the outer products of $M$ vectors.
If $M\ge T$ and the vectors $\xv^t$ are appropriately set, $\Gm$ can be any $T\times T$ Gram matrix.
$\Zm$ and $\Km$ are functions of $\Gm$.
Thus, setting $M=T$ gives the same result as setting $M\rightarrow\infty$.

The states of the input neurons $u_i^t$ are constants,
and the activities of the linear neurons $x_i^t$ and output neurons $y_i^t$ are unconstrained and constrained variables, respectively.
The states of the nonlinear neurons $v_i^t$ and connection weights are implicitly included in the kernel function.

\subsection{Formalization with recurrent neural network kernel}
In the numerical simulations, $\Km$ is replaced with $\Km+T\mu \Em$.
This modification is needed for three reasons.
First, it numerically stabilizes the inversion of $\Km$.
Second, it allows the inversion of $\Km$ in repetitive tasks (see Appendix~\ref{appendixcirculant}).
Third,
it accelerates the optimization.

To simplify the model, we set $\sigma_0$ and $\sigma_b$ to infinity while keeping $\sigma_b/\sigma_0=\beta$.
Thus, the kernel in the numerical experiments is given by

\begin{equation}
K_{ij} = \frac{2}{\pi}\arcsin\left(\frac{Z_{ij}}{
\sqrt{Z_{ii} Z_{jj}}
}\right)+\mu T\delta_{ij},
\end{equation}
where $\delta_{ij}=1$ if $i=j$ and $\delta_{ij}=0$ otherwise,
and $\Fm$ is replaced with $\Fm+\beta\onev\trans\onev$.
The use of finite $\sigma_0$ and $\sigma_b$ does not qualitatively change the simulation results, but it slows the optimization and obscures the results in many cases.

\subsection{Numerical simulation}
The numerical simulation is formulated as an inequality-constrained optimization problem that minimizes $\Lambda$ given variables $x_i^t$ and $y_i^t$.
The input $u_i^t$ is fixed at the beginning of the simulation.
In the constrained trials, the value of $y_i^t$ is constrained by inequalities to satisfy the task requirements.
In the unconstrained trials, $y_i^t$ is a free parameter.
The optimization is performed by Optizelle (\url{http://www.optimojoe.com/products/optizelle}).
The gradient and Hessian-vector product are derived in Appendix~\ref{Diff}.
Each simulation started the optimizations from 10 different initial values and presented the result with the smallest $\Lambda$.
The model parameter values were set to $\alpha=10^{-6}$, $\beta=1$ and $\mu=0.001$.
In Optizelle, the parameter values were set to algorithm\_class=``TrustRegion'', iter\_max=10000, delta=1, krylov\_solver=ConjugateDirection, eps\_krylov=0.01, krylov\_iter\_max=20, krylov\_orthog\_max=20, eps\_dx=$10^{-16}$, sigma=0.995, and gamma=0.7.
After 20 iterations, krylov\_iter\_max was set to 1000.

\section{Examples}
Here, we present the simulation results of the AND, delayed response, and motor-pattern generation tasks.

\subsection{AND task}
The AND task involves four input neurons and one output neuron.
The variables $u_1^t$, $u_2^t$, $u_3^t$, and $u_4^t$ are set to one in the time steps presenting the Cue, Input A, Input B, and Go signal, respectively; in other time steps, they are set to zero.
Cue marks the beginning of one trial of this task.
At that time, Inputs A and B are presented independently with probability 50\%.
The Go signal is presented 11 steps later.
The step following Go requires the inequality constraint $y_1^t\ge 1$ if both inputs are presented in the trial, and $y_1^t\le 0$ otherwise.
Other steps require the inequality constraint $y_1^t\le 0$.
To summarize, this task performs a delayed AND operation.
Each trial is followed by an intertrial delay period.
The whole task completes 40 sets of the 13-time-step trial and the 7-time-step intertrial delay period.
The last six (unconstrained) trials test whether the network can learn the task, that is, perform the task correctly without constraints.

\begin{figure}
  \centering
  \includegraphics{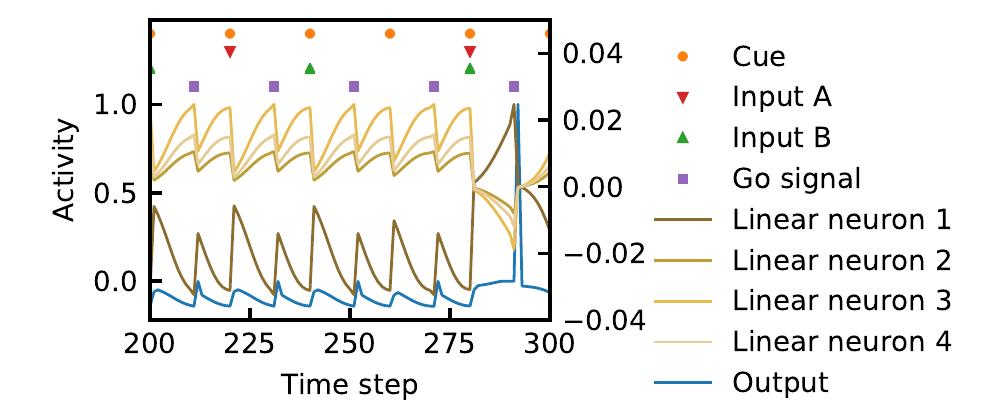}
    \caption{\label{ANDsustained}
      Symbols indicate the time steps of Cue (orange circle), Input A (red inverted triangles), Input B (green upright triangles), and Go signal (purple squares).
      The activity time courses of the output and linear neurons are indicated by the left and right axes, respectively.
    }
\end{figure}

Figure~\ref{ANDsustained} shows the activity of the linear and output neurons in the optimized model network.
When Inputs A and B are presented in the trial, the activity of the output neuron reaches one just after the Go signal; at other time steps, it is suppressed.
Therefore, the inequality constraints for $y_i^t$ are satisfied.
The linear neurons presented in Figure~\ref{ANDsustained} exhibit sustained or ramping activity depending on the value of A AND B.
One of the neurons increases its activity after the presentation of A and B; the remaining three decrease their activities.

\begin{figure}
  \centering
  \includegraphics{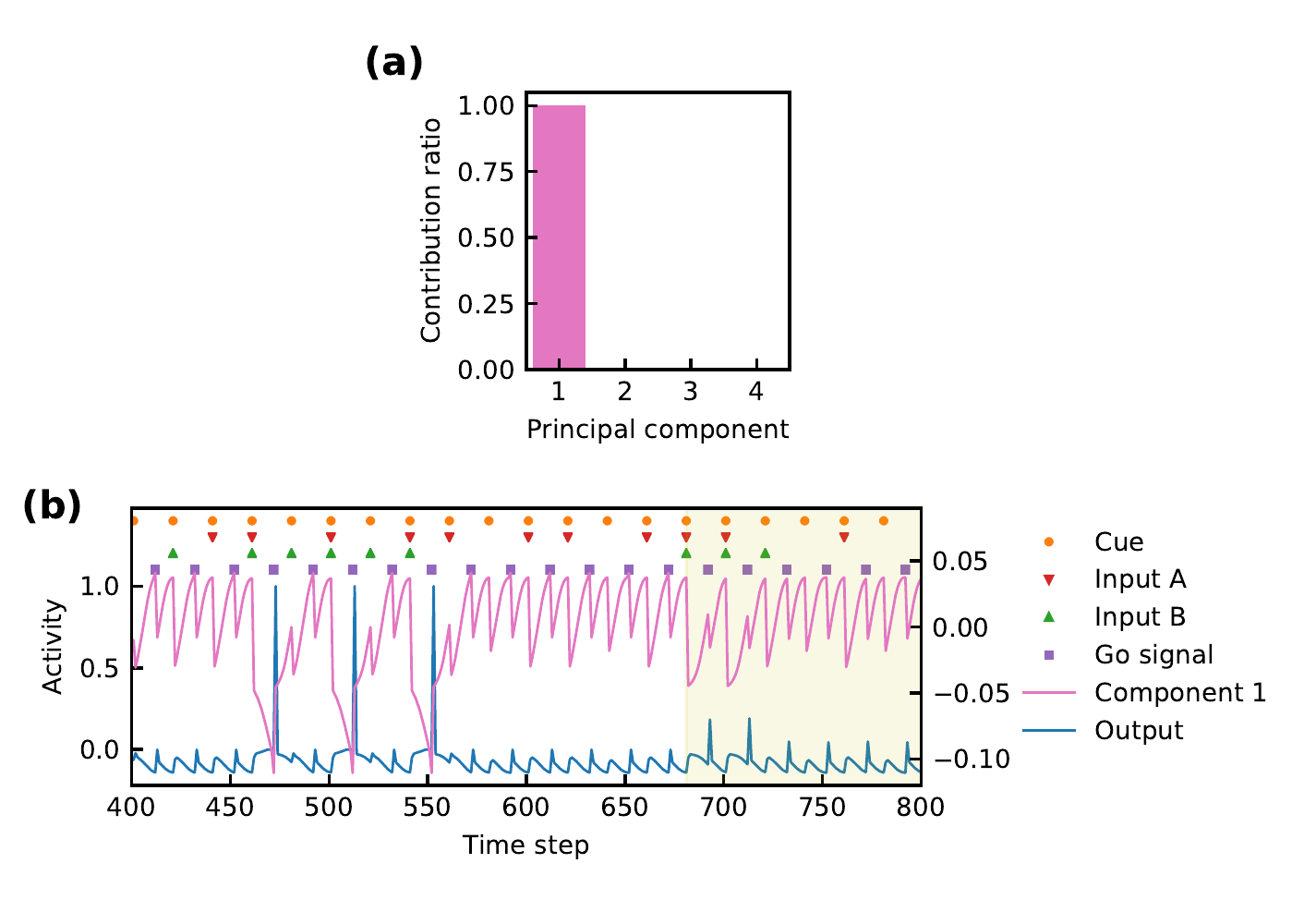}
    \caption{\label{AND}
      PCA of the AND task.
      (a) Contribution ratios of the four largest principal components.
      (b) Time courses of Cue, Input A, Input B, Go signal, output, and the largest component.
      The activities of the output neuron and the principal component are indicated by the left and right axes, respectively.
      The last six trials (shaded area) are unconstrained trials to test whether the network performs the task correctly.
    }
\end{figure}

The ramping activity exemplified in Figure~\ref{ANDsustained} is observed in working memory experiments \citep{Rowe2000}.
To demonstrate the prevalence of ramping activity among linear neurons, we apply principal component analysis (PCA).
Let us assume that the time course of the activity of neuron $i$ is given by the linear superposition of its time-varying components, that is,

\begin{equation}
x_i^t = \bar{x}_i+\sum_{c=1}^C w_{ic} e_c^t,
\end{equation}
where $\bar{x}_i$ is a constant, $e_c^t$ is the value of component $c$ at time step $t$, $w_{ic}$ is the weight of component $c$ on linear neuron $i$, and $C$ is the number of components.
The components are assumed to be orthonormal, that is,

\begin{equation}
\sum_{t=1}^T e_c^t e_{c'}^t=\delta_{cc'}.
\end{equation}
In PCA, each loading vector $\ev_c$ is an eigenvector of

\begin{equation}
(\Gm_{\mathrm{centered}})_{tt'} = \frac{1}{\sum_{t=1}^T\sum_{m=1}^M (x_m^t-\bar{x}_m)^2}\sum_{m=1}^M (x_m^t-\bar{x}_m) (x_m^{t'}-\bar{x}_m),
\end{equation}
where $\bar{x}_m=\frac{1}{T}\sum_t x_m^t$.
The eigenvalue corresponding to $\ev_c$ is given by $\sum_{i=1}^T w_{ic}^2$.
The contribution ratios of the four largest eigenvalues are shown in Figure~\ref{AND}a.
Most of the activity of the linear neurons is explained by the first component, $\ev_1$.
In other words, the time courses of the linear neurons are closely approximated by $x_m^t = \bar{x}_m+w_{m1} e_1^t$.
Figure~\ref{AND}b shows the time course of $\ev_1$, which stores the memory of A AND B.
In the last six (unconstrained) trials, the activities of the output neuron and the first principal component distinguish the value of A AND B.

In the present simulation, the value of A AND B is encoded and memorized by ramping or sustained activity.
Although this value can also be encoded by a synfire chain-like sequence of activity \citep{Abeles1991}, the optimization does not yield the sequence of activities.
This can be explained as follows.
Let us partition $N_0$ neurons into $N_G$ groups and assume that group $i$ activates group $i+1$.
Moreover, a neuron must receive a collective input $W_{\mathrm{bound}}$ to be activated.
If the activity is sustained, that is, $N_G=1$, each neuron in the group of $N_0$ neurons receives an input $W_{\mathrm{bound}}$.
If each neuron outputs one during sustained activity, it receives a total connection weight of $W_{\mathrm{bound}}$ from the neurons in the group.
Therefore, the approximated log-probability is proportional to $-W_{\mathrm{bound}}^2$.
On the other hand, if $N_G>1$ groups with $N_0/N_G$ neurons form a chain-like structure, the approximated log-probability is proportional to $-(N_G-1)W_{\mathrm{bound}}^2$ [see Equation~(\ref{probability1})].
Hence, sustained activity is more probable than a sequence of activity.
Note also that sustained activity is more robust to changes in the delay-period duration (i.e., is more generalizable) than a synfire chain-like activity.
Varying the durations of the delay period does not affect the overall dynamics of the optimized model (data not shown).



\subsection{Delayed response task}
\begin{figure}
  \centering
  \includegraphics{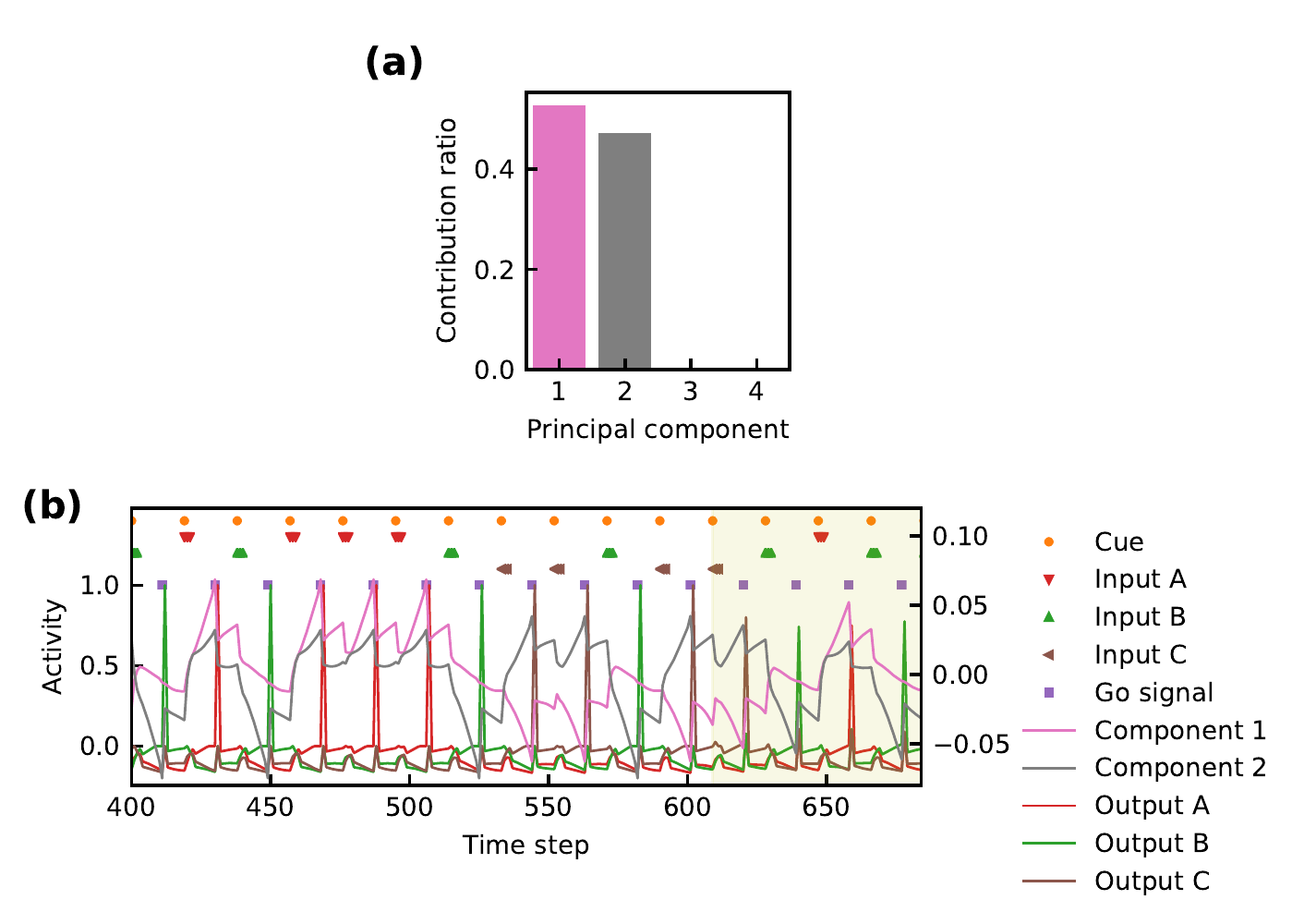}
    \caption{\label{recall}
      Dynamics of the delayed response task.
      (a) Contribution ratios of the four largest principal components.
      (b) Time courses of Cue, Input A, Input B, Input C, and the Go signal, the three outputs, and the two largest PCA components.
      $y_1$, $y_2$, and $y_3$ are indicated by Outputs A, B, and C, respectively.
      The activities of the output neurons and the principal components are indicated by the left and right axes, respectively.
      The last four trials (shaded area) are unconstrained trials to test whether the network performs the task correctly.
    }
\end{figure}

The second task is a delayed response task with three stimulus--response pairs.
The task involves five input neurons and three output neurons.
At the presentation times of Cue, Input A, Input B, Input C, and Go, $u_1^t$, $u_2^t$, $u_3^t$, $u_4^t$, and $u_5^t$ are set to one, respectively; at other time steps, they are set to zero.
As before, Cue marks the beginning of a trial.
At the same time, one of Input A, Input B, and Input C is presented and is continued over three time steps.
After a delay of eight time steps, the Go signal is presented.
The trials presenting Input A, Input B, and Input C require the inequality constraints $y_i^t\ge 1$ and $y_j^t\le 0$ ($j\neq i$) for $i=1$, 2, and 3, respectively, at the time step following the Go signal, and the constraint $y_i^t\le 0$ ($i=1$, 2, and 3) at other time steps.
Figure~\ref{recall} shows the simulation result.
Among the 36 trials, the last 4 trials are unconstrained trials.
The intertrial interval (between the Go signal and the next Cue) is six time steps.

In the PCA, the neuronal activity is represented as the linear superposition of two components (Figure~\ref{recall}a).
The first component increases and decreases after the presentation of Inputs A and C, respectively, whereas the second component increases after the presentation of both Inputs A and C and decreases after Input B (Figure~\ref{recall}b).
Although three stimuli must be memorized, the activity is explained by only two components, because three points on a plane are linearly separable.
Given that the approximated log-probabilities of three self-activating groups and two self-activating groups are proportional to $-3W_{\mathrm{bound}}^2$ and $-2W_{\mathrm{bound}}^2$, respectively, and hence the activity of linear neurons is a linear superposition of two components.
The model network outputs the correct activity in the last four unconstrained trials.

\subsection{Motor-pattern generation task}
The third task is a simplified version of the task reported in \cite{Churchland2012}.
In contrast to the previous tasks with binary outputs, the network in this task must output continuous motor patterns.
The variables $u_1^t$, $u_2^t$, $u_3^t$, and $u_4^t$ are set to one at the presentation times of Input A, Input B, Input C, and Go, respectively, and to zero at other time steps.
These three inputs, which are presented in three time steps, indicate the three motor patterns to be executed after the Go signal.
The duration $\tau$ of the motor patterns is 12 time steps.
This model has two outputs: $y_1^t$ and $y_2^t$.
The constraints on pattern $i$ are $|y_1^t-\sin(2\pi \phi_i(t-t_0)/\tau)|\le\Delta y$ and $|y_2^t - \sin(2\pi \psi_i(t-t_0)/\tau)|\le\Delta y$, where $t_0$ is the next time step of the Go signal, and $\Delta y=0.1$.
The frequencies $(\phi_i,\;\psi_i)$ are defined by $(\phi_1,\;\psi_1)=(1,\;0)$, $(\phi_2,\;\psi_2)=(0,\;1)$, and $(\phi_3,\;\psi_3)=(2,\;2)$.
During other time steps, the output constraints are $|y_1|\le\Delta y$ and $|y_2|\le\Delta y$.
The intertrial interval is eight time steps.
The simulation contains 36 trials: 30 constrained trials and 6 unconstrained trials.

\begin{figure}
  \centering
  \includegraphics{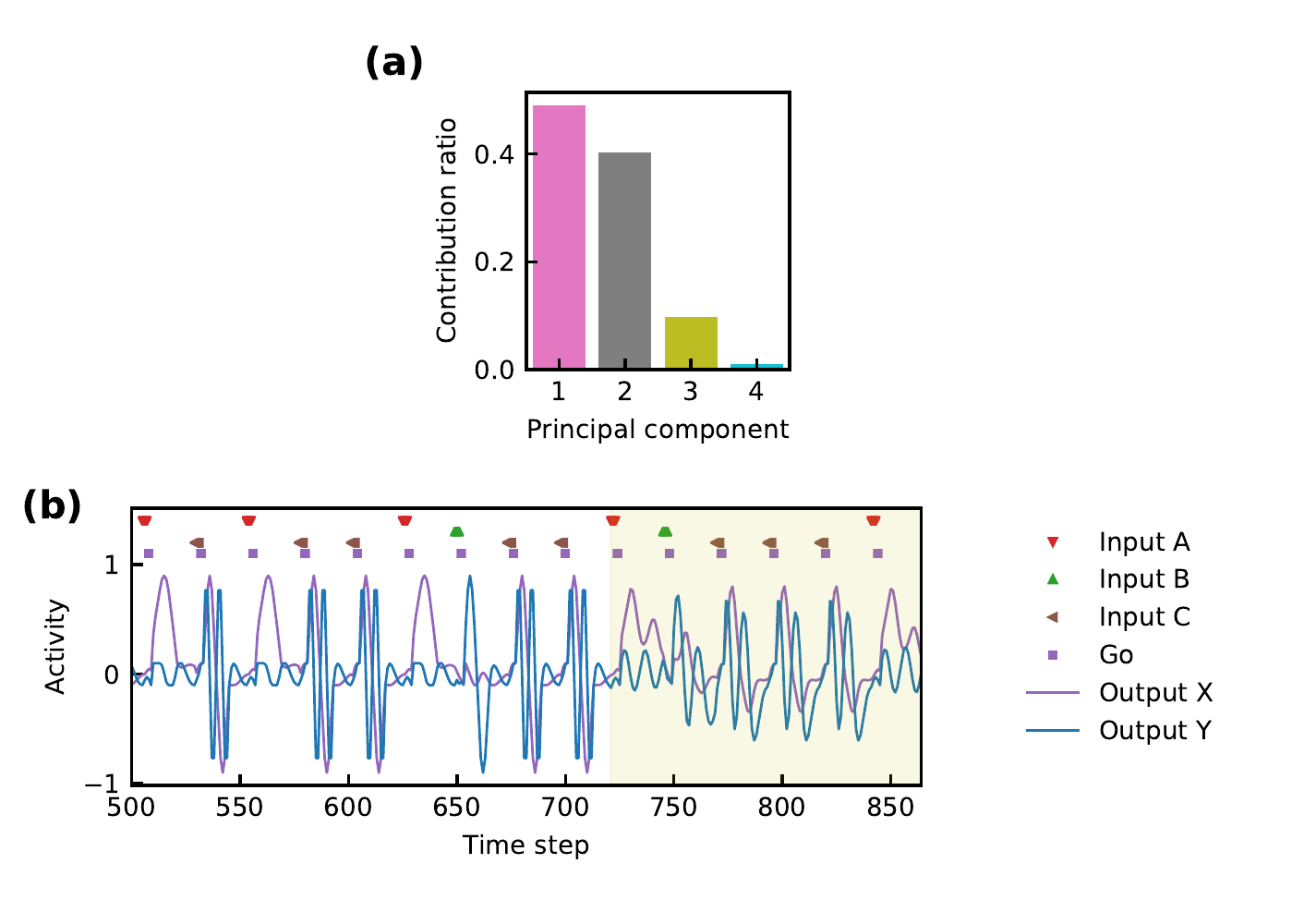}
    \caption{\label{movement}
      Motor-pattern generation task.
      (a) Contribution ratios of the four largest principal components.
      (b) Time courses of Input A, Input B, Input C, and the Go signal and the two outputs.
      The last six trials (shaded area) are unconstrained trials to test whether the network performs the task correctly.
    }
\end{figure}

\begin{figure}
  \centering
  \includegraphics{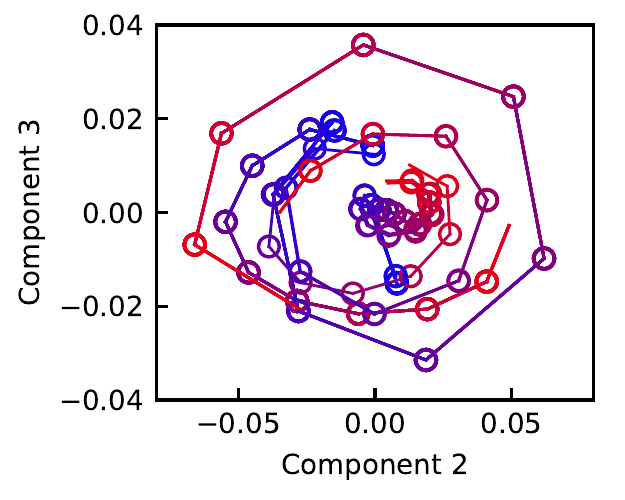}
    \caption{\label{movementtrajectory}
      Trajectories of the second and third largest components in the motor-pattern generation.
      Red and blue indicate the beginning and end of the motor patterns, respectively.
    }
\end{figure}

Figure~\ref{movement}a shows the contribution ratios of the four largest principal components, and Figure~\ref{movement}b shows the output and principal components of the optimized model for this task.
Owing to the complexity of this task, the number of relevant components in this task is larger than in the previous tasks.
The trajectories of the second and third largest components in the motor pattern generation are shown in Figure~\ref{movementtrajectory}.
The trajectories generally rotate clockwise despite the large differences among the three motor patterns.

Rotating PCA components were also reported by \cite{Churchland2012}.
Whereas \cite{Churchland2012} plotted the largest and second largest components, whereas we plotted the second and third largest components because the largest component rapidly changes similar to the outputs.
Both in our study and in \cite{Churchland2012}, different motor patterns were realized by slightly modulated activity trajectories.
This suggests that patterns realized by modulated trajectories require a lower total connection weight than those realized by completely different trajectories.
To construct two different trajectories, the total connection weight must be doubled; consequently, the circuit for two different trajectories is less probable.

\section{Discussion}
\subsection{Summary of the model and the results}
The present model predicts the activity of the most probable network in the limit of infinitely many neurons.
When optimized for the AND, delayed response, and motor-pattern generation tasks, the neuronal activity in the networks was a linear superposition of one, two, and four components, respectively.
Thus, we can regard the network as a mixture of groups of neurons.
Generally, the most probable network is a simple network of a few neuron groups.
In the motor-pattern generation task, the dynamics of the network components followed simple circular motions.

{The results of the present model are similar to those of \cite{Hennequin2014} and \cite{Sussillo2015}.
Similar to the present model, the previous models explained the simple dynamics of motor cortical neurons performing motor-pattern generation tasks.
However, the simple dynamics emerged from the stability optimization by optimizing the eigenvalues or minimizing the response to perturbation.
By contrast, the objective function of the present model contains no explicit terms for stabilizing or simplifying the dynamics.
Instead, the simple dynamics result from the maximization of the circuit probability.
This suggests that the simple dynamics in the brain are a byproduct of the most probable circuit.
}

The present model considers the activity and structure of networks and ignores the learning rule.
For a given task, it extracts the most probable activity and structure from the set of capable networks for the task.
Despite accumulating evidence for plasticity rules in the brain, the network is realized by these learning rules remains unclear.
However, as theoretically shown in Section 2, most of the capable networks for a task are concentrated around the most probable network.
Therefore, we can safely assume that the network realized by the brain's learning rule approximates the most probable network.

According to the analytical and simulation results, sustained activity is more favorable than an activity sequence.
That is, smaller connection weights among fewer groups are favored over greater connection weights among many groups.
Moreover, uniformly distributed connection weights are favored over nonuniformly distributed weights [see Equation~(\ref{con2})].
For these reasons, neurons with uniform connection weights tend to group together, and their sustained activity better accomplishes a task than an output sequence from several groups.
In this sense, the network structures formed by brain neurons and computer programs are fundamentally different.

\subsection{Relation to the algorithmic information theory}

Our hypothesis predicts that the brain finds the simplest and most probable network, thus implying that brain networks operate similarly to 
the simplest computer programs.
Therefore, our present hypothesis is analogous to algorithmic information theory \citep{Solomonoff1964a,Solomonoff1964b,Kolmogorov1965,Chaitin1966}, which measures the complexity of a sequence by the average length of the program that generates the sequence.
{Given that most of the randomly generated programs that can generate the sequence are the shortest programs, the average length is approximated by the shortest length.}
The average length is related to the probability that a randomly generated program will produce that sequence.
Similarly, the probability of task completion by a randomly generated network might measure the complexity of the task.

\begin{figure}
  \centering
  \includegraphics{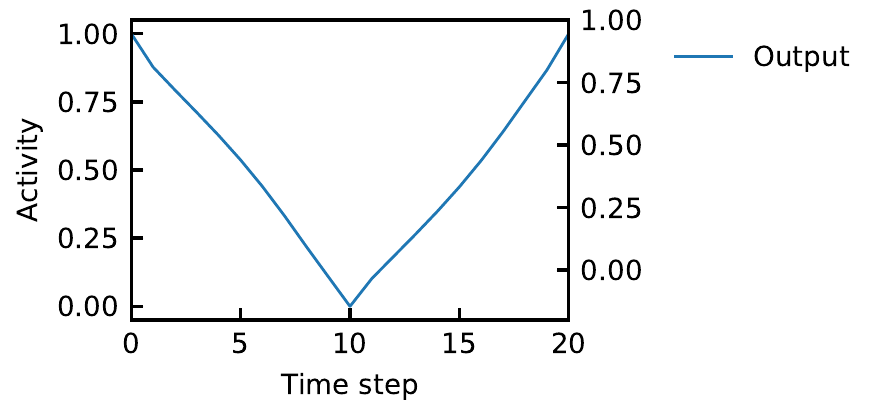}
    \caption{\label{checkpoint}
      Output of a model with no input and one output under the constraints $y_1^{10}\le 0$ and $y_1^{20}\ge 1$.
    }
\end{figure}

In algorithmic information theory, the shortest program naturally generalizes the sequence.
For example, consider the shortest computer program that generates one million digits of $\pi$ \citep{Cover2006},
possibly by a given formula.
Such a program will sequentially generate another million digits of $\pi$.
The most probable network (with the lowest connection weights) should be generalizable in the same manner.
{Considering that most of the networks that can perform a task are concentrated around the most probable network, they should be also generalizable.}
Networks with sustained activity can perform a task with a long or short delay period.
Under the present hypothesis, networks with low connection weights exhibit human-like generalization.
For example, suppose that in the absence of an input, an output alternates between two values.
Figure~\ref{checkpoint} shows the output of the most probable network in a 20-time step task under the constraints $y_1^{10}\le 0$ and $y_1^{20}\ge 1$.
As shown in the figure, $y_1^t$ continually increases from $t=1$ to $t=10$, and then continually decreases from $t=10$ to $t=20$.
Of course, neural networks can generate more complex trajectories.
High-frequency oscillations are exhibited by the networks with large $\Lambda$.
However, the most probable network forms a simple network structure with simple dynamics.
This suggests that simple network structures can describe the generalizations made by animals; consequently, simple computer programs can reproduce the intuitive generalizations of animals and humans.


\subsection{Experimental predictions}
The present model gives experimentally testable predictions.
First, the dynamics of a specifically trained network are limited to low-dimensional spaces.
Such low-dimensional dynamics can be clearer in membrane potentials than in spikes because dendritic potentials can be regarded as linear units.
According to the tightly balanced network hypothesis \citep{Chalk2016}, the membrane potentials of neighboring neurons are strongly correlated.
Consistent with this hypothesis, few of the principal components in the present model have large eigenvalues; consequently, the linear neurons are strongly correlated.
The optimized model exhibits low-dimensional dynamics, but an unlearned network may not exhibit low-dimensional dynamics.
Second, the neuronal activities of animals trained for a task exhibit similar Gram matrices.
This finding does not imply similar neuronal activities in two animals.
Indeed, even if the signs of all $x_i^t$ are reversed, the Gram matrix $\Zm$ and $\Lambda$ are preserved.
Third, information encoding in the brain is dominated by sustained activity rather than activity  sequences.
Fourth, the generalization exhibited by animals is replicated by the simplest neural network model;
that is, $L^2$ regularization allows animal- and human-like generalization in neural networks.

\subsection{Limitations and future directions}
Although the present model explains the sustained activity in the brain, it has several limitations.
First, it cannot learn certain tasks.
For example, the proposed model fails when the AND operation is replaced with an XOR operation unless Inputs A and B persist over three time steps.
Second, the neurons in the model are rate-coded in discrete-time, which precludes the modeling of spike synchrony and phase locking.
Third, our model cannot distinguish between excitatory and inhibitory neurons nor distinguish cell types, layer structures, and specialized regions.
Therefore, it cannot replicate cell- and region-specific firing patterns.
In other words, a group in our model includes all neuron types with any connection weight to any other group.
In real neuronal networks, plastic synaptic change is more easily realized in some directions than in other directions \citep{Sadtler2014}, thus indicating that the model assumptions are oversimplified.
Fourth, assuming that an infinite number of neurons is incompatible with the inherent features of brain networks.
For example, our model cannot account for fluctuations, which play important roles in encoding information \citep{Murray2014}.
Fifth, the saddle-point method is based on the central limit theorem, whereas synaptic weights apparently obey a long-tailed distribution \citep{Song2005,Teramae2012}, thus potentially leading to the nonstationary and chaotic behavior of the networks in the brain.
Therefore, real brain networks may exhibit more complex dynamics than the present model network.
  More importantly, stochastic binary neurons may not be approximated by deterministic sigmoid neurons.
In particular, averaging the activities of many neurons might alter the model behavior from that of brain networks.
Sixth, the network structure formed in the brain may depend on factors other than maximum probability, such as minimum energy consumption, minimum wiring lengths, and maximum information transmission \citep{Lennie2003,Chklovskii2004}.
Sparse coding, which maximizes the information transfer while minimizing the number of spikes, may also explain some features that are unexplained by the present model \citep{Olshausen1996}.
For example, the delayed response task with three stimulus--response pairs yielded two PCA components.
If each stimulus were encoded by three components, the sparser activity might better replicate the results of some experiments.

The present model may be extended and improved in several ways.
By dividing neurons into excitatory and inhibitory groups and constraining the sign of the connection weights, we might reveal the role played by each neuron type.
If the sparse activity of the linear neurons is enforced in the objective function, the activity might better resemble the experimental activity.
Under the sparseness term, the simulated motor activity might resemble the Gabor function-like basis functions emerging in sparse representations of natural scenes \citep{Olshausen1996}.
The elements emerging from a model network performing several tasks simultaneously are also of interest.





\appendix
\section{$L^2$ norm of the connection weights}\label{L2norm}
Let us derive the $L^2$ norm of the weight vector $\wv$ of the regression problem that minimizes \citep{Bishop2006}

\begin{equation}
E = \frac{1}{2}\sum_{i=1}^n [\wv^T\phiv(\xv_i)-t_i]^2+\frac{\lambda}{2} |\wv|^2,
\end{equation}
where $t_i$ is the target variable.
$E$ is minimized by

\begin{equation}
  \wv = \Phim^T\av,
\end{equation}
where $\av=-\lambda^{-1}(\Phim\wv-\tv)$, $\Phim=(\phiv(\xv_1),\;\ldots,\;\phiv(\xv_n))^T$, and $\tv=[t_1,\;\ldots,\;t_n]$.
Inserting $\wv = \Phim^T\av$ into $E$ yields

\begin{equation}
E = \frac{1}{2}\av^T\Phim\Phim^T\Phim\Phim^T\av
-\av^T\Phim\Phim^T\tv
+\frac{1}{2}\tv^T\tv
+\frac{\lambda}{2}\av^T\Phim\Phim^T\av,
\end{equation}
which is minimized by

\begin{equation}
  \av = (\Km+\lambda \Em)^{-1}\tv,
\end{equation}
where

\begin{equation}
\Km = \Phim\Phim^T.
\end{equation}
In the limit $\lambda\rightarrow0$, the $L^2$ norm of the weight vector is obtained as

\begin{align}
\wv^T\wv =& \tv^T\Km^{-1}\Phim\Phim^T\Km^{-1}\tv \nonumber\\
=& \Tr(\Km^{-1}\tv\tv^T).
\end{align}
For multiple target variables, $\tv\tv^T$ is replaced by the Gram matrix of the variables.

\section{Inverse of a circulant matrix}\label{appendixcirculant}
Let us define

\begin{equation}
\Am = \begin{pmatrix}
\mu mn\Em+\Km & \Km & \cdots & \Km\\
\Km & \mu mn\Em+\Km & \cdots & \Km\\
\vdots & \vdots & \ddots & \vdots\\
\Km & \Km & \cdots & \mu mn\Em+\Km
\end{pmatrix},
\end{equation}
where $\Km$ is an $m\times m$ symmetric matrix.
Assuming that $\Km$ has eigenvectors $\uv_i\;(1\le i\le m)$ and corresponding eigenvalues $\lambda_i\;(1\le i\le m)$, the eigenvectors of $\Am$ is given by

\begin{equation}
\uv_{i,j} = \begin{bmatrix}
\uv_i\\
\omega_n^j\uv_i\\
\vdots\\
\omega_n^{(n-1)j}\uv_i
\end{bmatrix}
\;(1\le i\le m,\;1\le j\le n),
\end{equation}
where $\omega_n=\exp\left(\im\frac{2\pi}{n}\right)$.
The corresponding eigenvalues are

\begin{equation}
\lambda_{i,j} =
\begin{cases}
\mu mn & j<n\\
\mu mn+n\lambda_i & j=n
\end{cases}.
\end{equation}
Assuming $|\uv_i|^2=1$, we have $|\uv_{i,j}|^2=n$.
Thus, the inverse $\inv\Am$ is given by

\begin{equation}
\inv\Am = \sum_{i=1}^m\sum_{j=1}^n \lambda_{i,j}^{-1}\frac{1}{n}\uv_{i,j}\uv_{i,j}^\dagger.
\end{equation}
By defining

\begin{equation}
\gv = \begin{bmatrix}
\tilde\gv\\
\tilde\gv\\
\vdots\\
\tilde\gv
\end{bmatrix},
\end{equation}
we obtain

\begin{align}
\uv_{i,j}^\dagger\gv
= & \sum_{k=1}^{n} \omega_n^{-jk}\uv_i^\dagger\tilde\gv \nonumber\\
= & \begin{cases}
0 & j<n\\
n\uv_i^\dagger\tilde\gv & j=n
\end{cases}.
\end{align}
Hence, we have

\begin{equation}
\trans\gv\inv\Am\gv = \sum_{i=1}^m \frac{1}{\mu m+\lambda_i}|\uv_i^\dagger\tilde\gv|^2,
\end{equation}
from which it follows that
$\Tr(\inv\Am\Gm)$ of the Gram matrix $\Gm$ of is independent of $n$.
Here $\Gm$ comprises the vectors corresponding to an $m$-periodic activity.

\section{Differentiation of $\Lambda$}\label{Diff}
Before differentiating $\Lambda$, let us define the $\circ$ and $\bullet$ operators for $n\times n$ matrices $A$ and $B$ by

\begin{align}
(\Am\circ\Bm)_{ij} =& A_{ij}B_{ij},\\
\Am\bullet\Bm =& \sum_{i=1}^n \sum_{j=1}^n A_{ij}B_{ij},
\end{align}
the element-wise mapping and element-wise power operators by

\begin{align}
  (f[\Am])_{ij} &= f(A_{ij}),\\
  ([\Am]^a)_{ij} &= A_{ij}^a,
\end{align}
the time-shift operators $^+$ and $^-$ by

\begin{align}
  (\Am^+)_{ij} =&
    \begin{cases}
    A_{i+1\; j+1} & i<n,\; j<n\\
    A_{1\;j+1} & i=n,\; j<n\\
    A_{i+1\;1} & i<n,\; j=n\\
    A_{11} & i=n,\; j=n
  \end{cases},\\
(\Am^-)_{ij} =& \begin{cases}
    A_{i-1\; j-1} & i>1,\; j>1\\
    A_{n\;j-1} & i=1,\; j>1\\
    A_{i-1\;n} & i>1,\; j=1\\
    A_{nn} & i=1,\; j=1
  \end{cases},
\end{align}
and the diagonal matrix-vector operators $\diag$ by

\begin{gather}
  [\diag\Am]_i = A_{ii},\\
  (\diag\vv)_{ij} = \delta_{ij}v_i,
\end{gather}
where $\vv$ is an $n$-dimensional vector.
Finally, the variation is given by

\begin{equation}
(\vd_{\Bm} \Am)_{ij} = \sum_{k=1}^n\sum_{l=1}^n\pdiff{A_{ij}}{B_{kl}}\delta B_{kl}.
\end{equation}
The element-wise power $[\Am]^{-1}$ should not be confused with the matrix inverse $\Am^{-1}$.
The $\bullet$ notation is simplified as

\begin{equation}
\Am^{(1)}\bullet\Am^{(2)}\bullet\cdots\bullet\Am^{(m)} = \sum_{i=1}^n\sum_{j=1}^n\prod_{k=1}^m A^{(k)}_{ij}.
\end{equation}
The following formulas can be easily proven:

\begin{gather}
\Am\circ\Bm = \Bm\circ\Am,\\
\Am\bullet\Bm = \Bm\bullet\Am,\\
\Am\circ(\Bm\circ\Cm) = (\Am\circ\Bm)\circ\Cm,\\
\Am\bullet(\Bm\circ\Cm) = \Am\bullet\Bm\bullet\Cm,\\
(\Am\Bm\Cm)\bullet\Dm = \Bm\bullet(\trans\Am\Dm\trans\Cm),\\
\Am^\pm\bullet\Bm = \Am\bullet\Bm^\mp,\\
f[\Am^\pm] = (f[\Am])^\pm,\\
(\diag\Am)\bullet\vv = \Am\bullet \diag\vv,\\
\vd_{\Cm}(\Am\circ\Bm) = \vd_{\Cm}\Am\circ\Bm+\Am\circ\vd_{\Cm}\Bm,\\
\vd_{\Cm}(\Am\bullet\Bm) = \vd_{\Cm}\Am\bullet\Bm+\Am\bullet\vd_{\Cm}\Bm,\\
\vd_{\Cm}(\Am\Bm) = (\vd_{\Cm}\Am)\Bm+\Am(\vd_{\Cm}\Bm),\\
\vd_{\Bm} f[\Am] = f'[\Am]\circ\vd_{\Bm}\Am,\\
\vd_{\Bm}(\Am^\pm) = (\vd_{\Bm}\Am)^\pm,\\
\vd_{\Bm}\Am^{-1} = -\Am^{-1}\vd_{\Bm}\Am\Am^{-1}.
\end{gather}
Note that $\diag$, $\vd_{\Am}$, and $^\pm$ are linear operators, whereas $\circ$ and $\bullet$ are bilinear operators.
By definition, we obtain

\begin{equation}
  (\vd_{\Am} \Am)_{ij} = \delta A_{ij} = (\vd\Am)_{ij}.
\end{equation}
The relation

\begin{equation}
  \vd_{\Am}\Lambda = \vd\Am\bullet\Bm,
\end{equation}
implies

\begin{equation}
  \pdiff{\Lambda}{A_{ij}} = B_{ij}.
\end{equation}

By defining the $T\times M$ matrix $\Xm=(x_m^t)$, the $T\times L$ matrix $\Ym=(y_l^t)$, and the $T\times K$ matrix $\Um=(u_k^t)$, the Gram matrices $\Gm$, $\Hm$, $\Fm$, and $\Zm$ are written as follows:

\begin{gather}
\Gm = \Xm\trans\Xm,\\
\Hm = \Ym\trans\Ym,\\
\Fm = \Um\trans\Um+\beta\Em,\\
\Zm = \Fm+\Gm.
\end{gather}
From the definition

\begin{gather}
\Lambda = \inv\Km\bullet(\Gm^++\Hm^+),\\
\Km = \frac{2}{\pi}\arcsin\left[\Zm\circ[\zv\trans\zv]^{-1/2}\right],\\
\zv = \diag\Zm,
\end{gather}
we obtain

\begin{align}
\vd_{\Gm}\Km =& \frac{2}{\pi}\left[1-\left[\Zm\circ[\zv\trans\zv]^{-1/2}\right]^2\right]^{-1/2} \nonumber\\
& \circ\left(\vd\Gm\circ[\zv\trans\zv]^{-1/2}-\Zm\circ\frac{1}{2}[\zv\trans\zv]^{-3/2}\circ (\diag[\vd\Gm]\trans\zv+\zv\trans{\diag[\vd\Gm]})\right) \nonumber\\
=& \Rm \circ\left(2\vd\Gm-\Zm\circ[\zv\trans\zv]^{-1}\circ (\diag[\vd\Gm]\trans\zv+\zv\trans{\diag[\vd\Gm]})\right),
\end{align}
where

\begin{equation}
\Rm = \frac{1}{\pi}\left[\zv\trans\zv-\Zm\circ\Zm\right]^{-1/2}.
\end{equation}
Considering that $Z_{ii}/z_i=1$, we can set $K_{ii}=1$ and $R_{ii}=0$.
Furthermore, if $z_iz_j-Z_{ij}^2$ ($i\neq j$) is below zero because of numerical error, it can be is assumed as $10^{-5}$.
Thus we have

\begin{align}
\vd_{\Gm}\Lambda =& -(\inv\Km\vd_{\Gm}\Km\inv\Km)\bullet(\Gm^++\Hm^+)
+\inv\Km\bullet\vd_{\Gm}\Gm^+ \nonumber\\
=& \vd\Gm\bullet[-2\Rm\circ[\inv\Km(\Gm^++\Hm^+)\inv\Km]+(\inv\Km)^-] \nonumber\\
&+2\diag[\vd\Gm]\bullet[(\Rm\circ\Zm \circ[\zv\trans\zv]^{-1}\circ[\inv\Km(\Gm^++\Hm^+)\inv\Km])\zv] \nonumber\\
=& \vd\Gm\bullet[-\Rm\circ\Am+(\inv\Km)^-]
+\diag[\vd\Gm]\bullet[(\Rm\circ\Zm \circ[\zv\trans\onev]^{-1}\circ\Am)\onev],
\end{align}
where

\begin{equation}
\Am = 2\inv\Km(\Gm^++\Hm^+)\inv\Km.
\end{equation}
By replacing $\delta\Gm$ with $\delta\Gm'$ and differentiating again with respect to $\Gm$, we obtain

\begin{align}
\vd_{\Gm}\vd_{\Gm'}\Lambda
 =& \vd\Gm'\bullet\{-\vd_{\Gm}\Rm\circ\Am
-\Rm\circ\vd_{\Gm}\Am-(\inv\Km\vd_{\Gm}\Km\inv\Km)^-\} \nonumber\\
&+\diag[\vd\Gm']\bullet\{
(\vd_{\Gm}\Rm\circ\Zm \circ[\zv\trans\onev]^{-1}\circ\Am
+\Rm\circ\vd\Gm \circ[\zv\trans\onev]^{-1}\circ\Am \nonumber\\
&-\Rm\circ\Zm \circ[\zv\trans\onev]^{-2}\circ(\vd_{\Gm}\zv\trans\onev)\circ\Am
+\Rm\circ\Zm \circ[\zv\trans\onev]^{-1}\circ\vd_{\Gm}\Am)\onev\},
\end{align}
where

\begin{align}
\vd_{\Gm}\Rm =& -\frac{1}{2\pi}\left[\zv\trans\zv-\Zm\circ\Zm\right]^{-3/2}
\circ\left(\vd_{\Gm}\zv\trans\zv+\zv\trans{\vd_{\Gm}\zv}-2\vd\Gm\circ\Zm\right),\\
\vd_{\Gm}\zv =& \diag[\vd\Gm],\\
\vd_{\Gm}\Am =& 
-\inv\Km\vd_{\Gm}\Km\Am
-\Am\vd_{\Gm}\Km\inv\Km
+2\inv\Km\vd_{\Gm}\Gm^+\inv\Km.
\end{align}
Similarly, we obtain

\begin{equation}
\vd_{\Hm}\Lambda = \vd\Hm\bullet(\inv\Km)^-
\end{equation}
and

\begin{align}
\vd_{\Hm}\vd_{\Hm'}\Lambda =& \vd\Hm'\bullet(\vd_{\Hm}\inv\Km)^- \nonumber\\
=& 0,\\
\vd_{\Gm}\vd_{\Hm'}\Lambda =& \vd\Hm'\bullet(\vd_{\Gm}\inv\Km)^- \nonumber\\
 =& \vd\Hm'\bullet(-\inv\Km\vd_{\Gm}\Km\inv\Km)^-.
\end{align}
For two independent variables $\Xm$ and $\Ym$, we obtain

\begin{align}
\vd_{\Xm}\Lambda =&
2(\vd\Xm\trans\Xm)\bullet[-\Rm\circ\Am+(\inv\Km)^-]
+2\diag[\vd\Xm\trans\Xm]\bullet[(\Rm\circ\Zm \circ[\zv\trans\onev]^{-1}\circ\Am)\onev],\\
\vd_{\Ym}\Lambda =& 2(\vd\Ym\trans\Ym)\bullet(\inv\Km)^-,\\
\vd_{\Xm}\vd_{\Xm'}\Lambda
 =& 2(\vd\Xm'\trans\Xm)\bullet\{-\vd_{\Gm}\Rm\circ\Am
-\Rm\circ\vd_{\Gm}\Am-(\inv\Km\vd_{\Gm}\Km\inv\Km)^-\} \nonumber\\
&+2\diag[\vd\Xm'\trans\Xm]\bullet\{
(\vd_{\Gm}\Rm\circ\Zm \circ[\zv\trans\onev]^{-1}\circ\Am
+\Rm\circ\vd_{\Gm}\Zm \circ[\zv\trans\onev]^{-1}\circ\Am \nonumber\\
&-\Rm\circ\Zm \circ[\zv\trans\onev]^{-2}\circ(\vd_{\Gm}\zv\trans\onev)\circ\Am
+\Rm\circ\Zm \circ[\zv\trans\onev]^{-1}\circ\vd_{\Gm}\Am)\onev\} \nonumber\\
&+2(\vd\Xm'\trans{\vd\Xm})\bullet[-\Rm\circ\Am+(\inv\Km)^-]
+2\diag[\vd\Xm'\trans{\vd\Xm}]\bullet[(\Rm\circ\Zm \circ[\zv\trans\onev]^{-1}\circ\Am)\onev],\\
\vd_{\Ym}\vd_{\Ym'}\Lambda =& 2(\vd\Ym'\trans{\vd\Ym})\bullet(\inv\Km)^-,\\
\vd_{\Xm}\vd_{\Ym'}\Lambda =& 2(\vd\Ym'\trans\Ym)\bullet(-\inv\Km\vd_{\Gm}\Km\inv\Km)^-,\\
\vd_{\Ym}\vd_{\Xm'}\Lambda =&
4(\vd\Xm'\trans\Xm)\bullet[-\Rm\circ(\inv\Km\vd\Hm^+\inv\Km)] \nonumber\\
&+4\diag[\vd\Xm'\trans\Xm]\bullet[(\Rm\circ\Zm \circ[\zv\trans\onev]^{-1}\circ(\inv\Km\vd\Hm^+\inv\Km))\onev],
\end{align}
which give the gradient and Hessian-vector product needed for the optimization.

\section*{Conflict of Interest Statement}

The authors declare that the research was conducted in the absence of any commercial or financial relationships that could be construed as a potential conflict of interest.

\section*{Funding}
This work was supported by JSPS KAKENHI Grant Numbers 15H04266, 16K16123, and 16H04663.

\bibliographystyle{agsm}
\bibliography{manuscript}

\end{document}